\documentclass[final,1p,times]{elsarticle} 

\usepackage{graphicx}
\usepackage{amssymb} 
\usepackage{amsthm} 
\usepackage{lineno}

\journal{Nuclear Physics A} 

\begin{document}

\begin{frontmatter} 

\title{Collisional vs. Radiative Energy Loss of Heavy Quark in a Hot and Dense Nuclear Matter}

\author{Shanshan Cao, Guang-You Qin, Steffen A. Bass and Berndt M\"uller}
\address{Department of Physics, Duke University, Durham, North Carolina 27708, USA}


\begin{abstract} 
We study the heavy quark evolution in a quark-gluon plasma medium within the framework of Langevin equation coupled to a (2+1)-dimensional viscous hydrodynamic model. We modify the current Langevin algorithm such that apart from quasi-elastic scattering, medium-induced radiative energy loss is incorporated as well by treating gluon radiation as an extra force term. We find a significant effect of gluon radiation on heavy quark energy loss at LHC energies. Our calculation provides a good description of the D meson suppression measured by ALICE experiment, and makes a prediction for B meson suppression and flow.
\end{abstract} 

\end{frontmatter} 


\section{Introduction}

A highly excited deconfined state of QCD matter, the {\it strongly interacting quark-gluon plasma} (sQGP), is believed to be produced in relativistic heavy-ion collisions at RHIC and LHC. It displays properties similar to a nearly perfect fluid and has been successfully described by hydrodynamic models. Among various probes of the medium properties, heavy flavor has generated significant interest in the past decade, since heavy flavor high-$p_T$ suppression and elliptic flow  have been revealed as significant as for light quarks in spite of the large mass \cite{Adare:2010de,Bianchin:2011fa,Rossi:2011nx}, indicating a stronger coupling strength between heavy quark and QGP than expected. Therefore, it appears crucial to explore how heavy quarks interact with such a hot and dense medium.

Between the two energy loss mechanisms of heavy quark, gluon radiation is usually considered negligible compared with quasi-elastic scattering, as long as the heavy quark energy is not very large, because of the ``dead-cone effect" \cite{Dokshitzer:2001zm}. In the limit of multiple scatterings where the momentum transfer during each interaction is small, collisional energy loss inside a thermalized medium can be well described by the Langevin equation \cite{Moore:2004tg,He:2011qa}. In this framework, \cite{He:2011qa} provides a nice description of heavy flavor suppression and elliptic flow measured by RHIC. However, as we extend such study to the LHC energy level, even heavy quarks become ultra-relativistic, and therefore, it may no longer be reasonable to ignore the radiative energy loss due to the ``dead-cone effect". 

In this paper, we shall modify the current Langevin approach such that gluon radiation is also incorporated as an extra force term, whose momentum space distribution will be governed by the Higher-Twist calculation \cite{Guo:2000nz,Majumder:2009ge,Zhang:2003wk}. Within this improved approach, the competition between the two energy loss mechanisms will be clearly displayed as the energy increases. Combined with a (2+1)-dimensional hydrodynamic simulation of the QGP medium \cite{Song:2007fn,Song:2007ux,Qiu:2011hf}, we find our calculation of D meson suppression to be consistent with the LHC data.

\section{Methodology}
We modify the Langevin equation as follow:
\begin{equation}
\frac{d\vec{p}}{dt}=-\eta_D(p)\vec{p}+\vec{\xi}+\vec{f_g}.
\end{equation}
The first two terms on the right are the drag force and the thermal random force from the original Langevin equation for Brownian motion, and the third term $\vec{f_g}=-d\vec{p_g}/dt$ is introduced for the force exerted on the heavy quark by gluon radiation. The probability of gluon radiation and the momentum distribution of the radiated gluon ($\Delta\vec{p_g}$) during each time interval $\Delta t$ is sampled according to Eq.(\ref{Wang}) taken from the Higher-Twist calculation \cite{Guo:2000nz,Majumder:2009ge,Zhang:2003wk}:
\begin{equation}
\label{Wang}
\frac{dN_g}{dx dk_\perp^2 dt}=\frac{2\alpha_s(k_\perp)}{\pi} P(x) \frac{\hat{q}}{k_\perp^4} \textnormal{sin}^2\left(\frac{t-t_i}{2\tau_f}\right)\left(\frac{k_\perp^2}{k_\perp^2+x^2 M^2}\right)^4,
\end{equation}
in which $\hat{q}$ is the gluon transport coefficient, $k_\perp$ is the transverse momentum of the radiated gluon, and $x$ is the ratio between the gluon energy $\omega$ and the heavy quark energy $E$. Additionally, $\tau_f$ represents the gluon formation time and $P(x)$ is the splitting function. Note that the last bracketed term in Eq.(\ref{Wang}) is the ``dead cone factor" for a heavy quark with mass $M$.

With the assumption that the interaction during each scattering is small, the fluctuation-dissipation relation between the drag and the thermal force still holds -- $\eta_D(p)=\kappa/(2TE)$, where $\kappa$ is the momentum space diffusion coefficient defined in $\langle\xi^i(t)\xi^j(t')\rangle=\kappa\delta^{ij}\delta(t-t')$. Meanwhile, we set the lower limit of the gluon energy to be $\omega_0=\pi T$, which is the balancing point between gluon emission and absorption. These assumptions guarantee the equilibrium of heavy quarks after a sufficiently long time of evolution in the medium. The different transport coefficients are related via $D=2T^2/\kappa$ and $\hat{q}=2\kappa C_A/C_F$, where $D$ is the spatial diffusion coefficient of the heavy quark in the QGP.

We shall use our modified Langevin equation to simulate the heavy quark evolution inside a QGP. The QGP medium is generated with a (2+1)-dimensional hydrodynamic model \cite{Song:2007fn,Song:2007ux,Qiu:2011hf} with the MC-Glauber initialization. And the coupling of the Langevin approach to the QGP is in the same way as our previous study \cite{Cao:2011et,Cao:2012jt}. Our heavy quarks are initialized with the MC-Glauber model for the position space and a leading-order pQCD calculation for the momentum space. After traversing the medium, they are fragmented to heavy mesons via Pythia 6.4 \cite{Sjostrand:2006za}.

\begin{figure}[htb]
\begin{minipage}{0.47\textwidth}
\begin{center}
\includegraphics[width=0.85\textwidth,clip=]{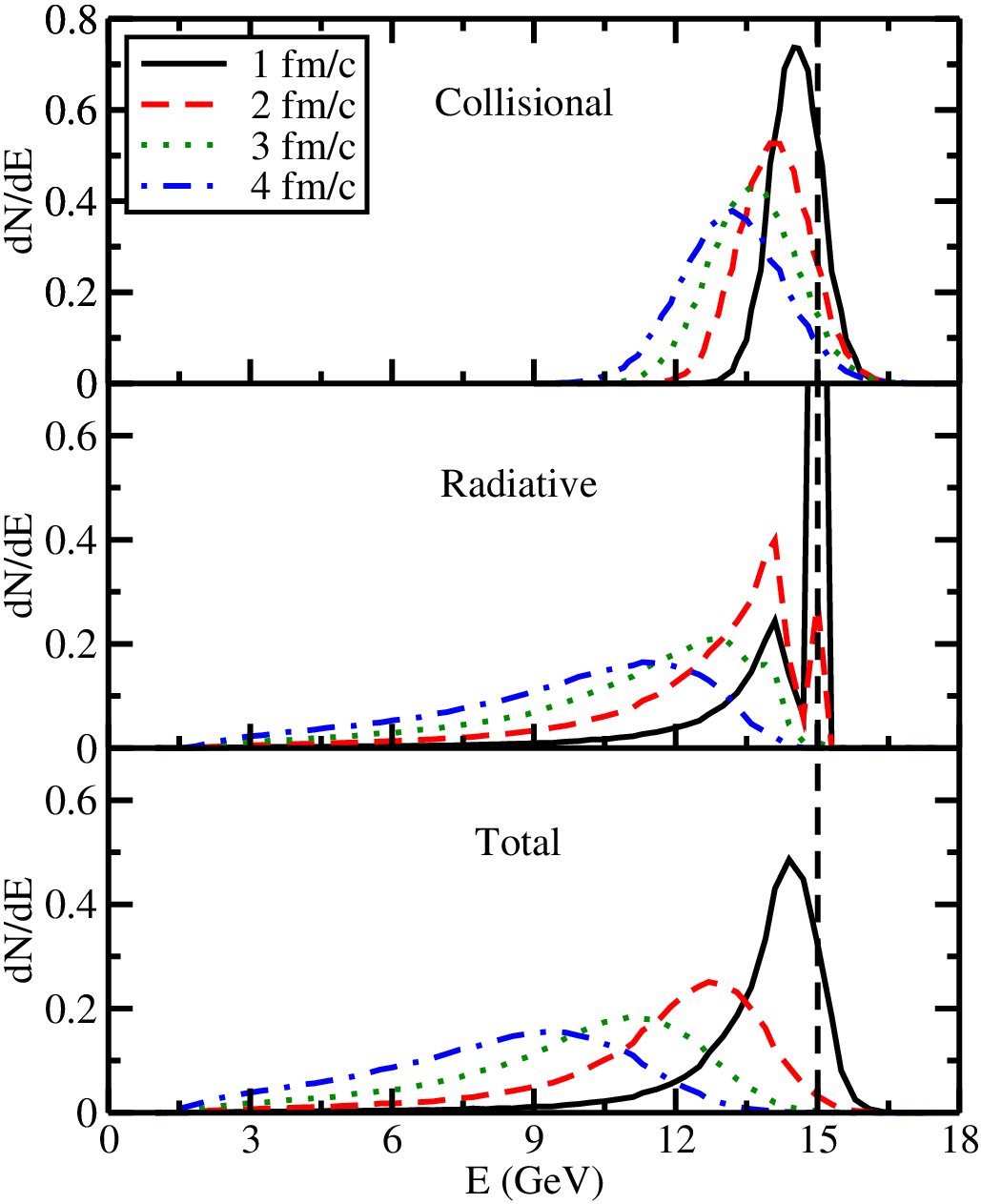}
\end{center}
\vspace{-12pt}
\caption{\label{Edistribution}(Color online) A comparison of the evolution of the charm quark energy distribution in a static medium between collisional, radiative and total energy loss.}
\end{minipage}\hspace{0.06\textwidth}
\begin{minipage}{0.47\textwidth}
\includegraphics[width=1\textwidth,clip=]{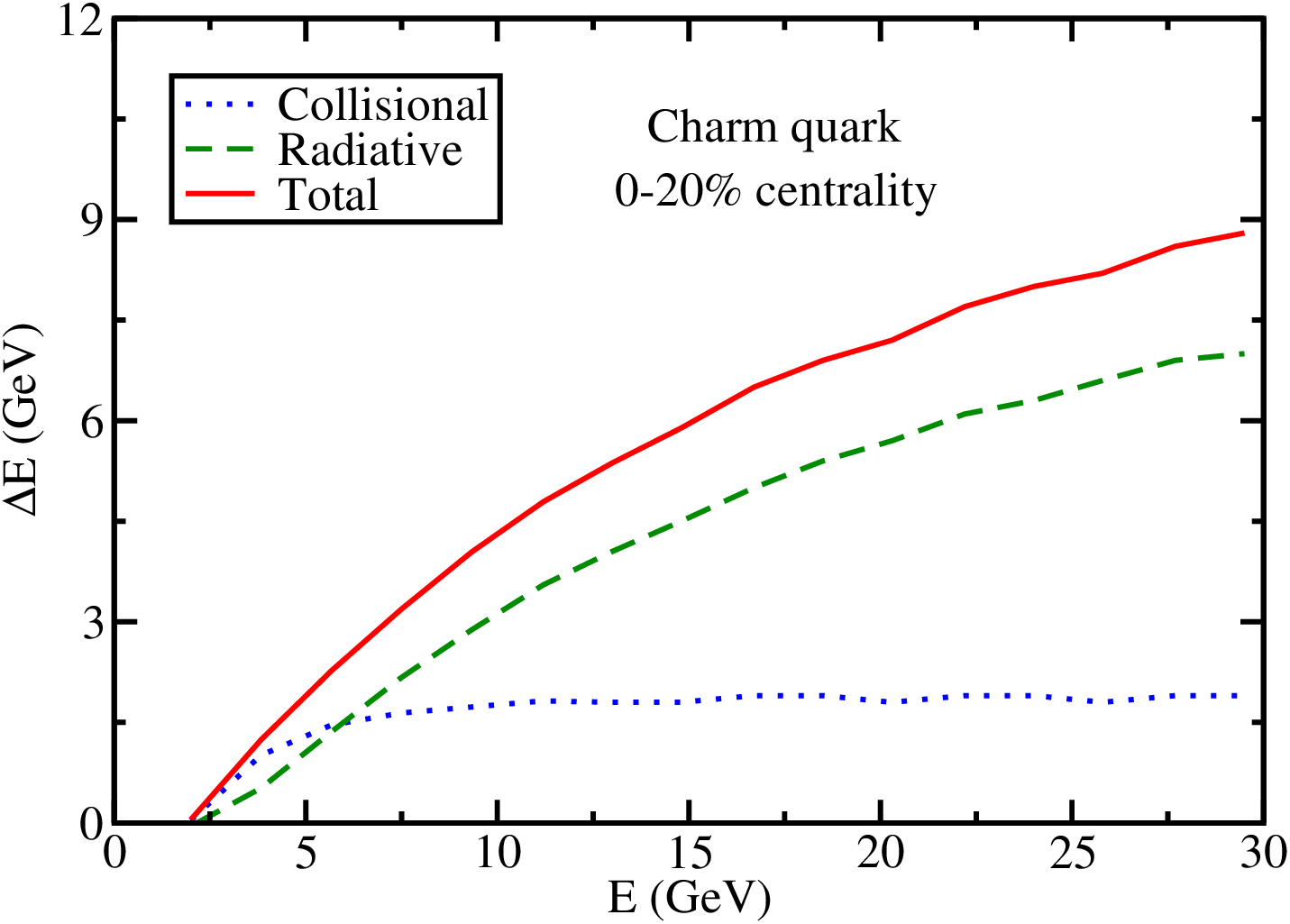}
\caption{\label{eloss}(Color online) A comparison of the charm quark energy loss after they traverse the QGP medium created in central Pb-Pb collisions with $\sqrt{s}=2.76$~TeV between collisional energy loss only, radiative only and the total effect. The crossing point between the collisional dominating region and the radiative dominating region is around 6~GeV.}
\end{minipage}
\end{figure}

\section{Heavy flavor evolution, suppression and flow}

We start with the evolution of the charm quark energy spectrum due to the different energy loss mechanisms. Here, the charm quarks are all initialized with 15~GeV energy before traveling through an infinite medium with a fixed temperature of 300~MeV. The spatial diffusion coefficient in our study is set at $D=6/(2\pi T)$, corresponding to a gluon transport coefficient $\hat{q}$ around 1.3~GeV$^2$/fm at 300~MeV. As shown by Fig.\ref{Edistribution}, before 2~fm/c, collisional energy loss dominates the charm quark evolution. However, after 2~fm/c, gluon radiation starts to dominate.

Fig.\ref{eloss} displays the energy loss of charm quarks after propagating through a realistic QGP medium produced in Pb-Pb collisions at the LHC energy level. The $x$-axis represents the initial energy of charm quarks and $y$ represents the energy loss. One observes that while collisional energy loss dominates the low energy region, gluon radiation dominates the high energy region. The crossing point is around 6~GeV, indicating that the collisional energy loss alone may describe the heavy flavor observed by the current RHIC measurement well, but will become insufficient when we extend to LHC energies.

\begin{figure}[htb]
\begin{minipage}{0.47\textwidth}
\includegraphics[width=1\textwidth,clip=]{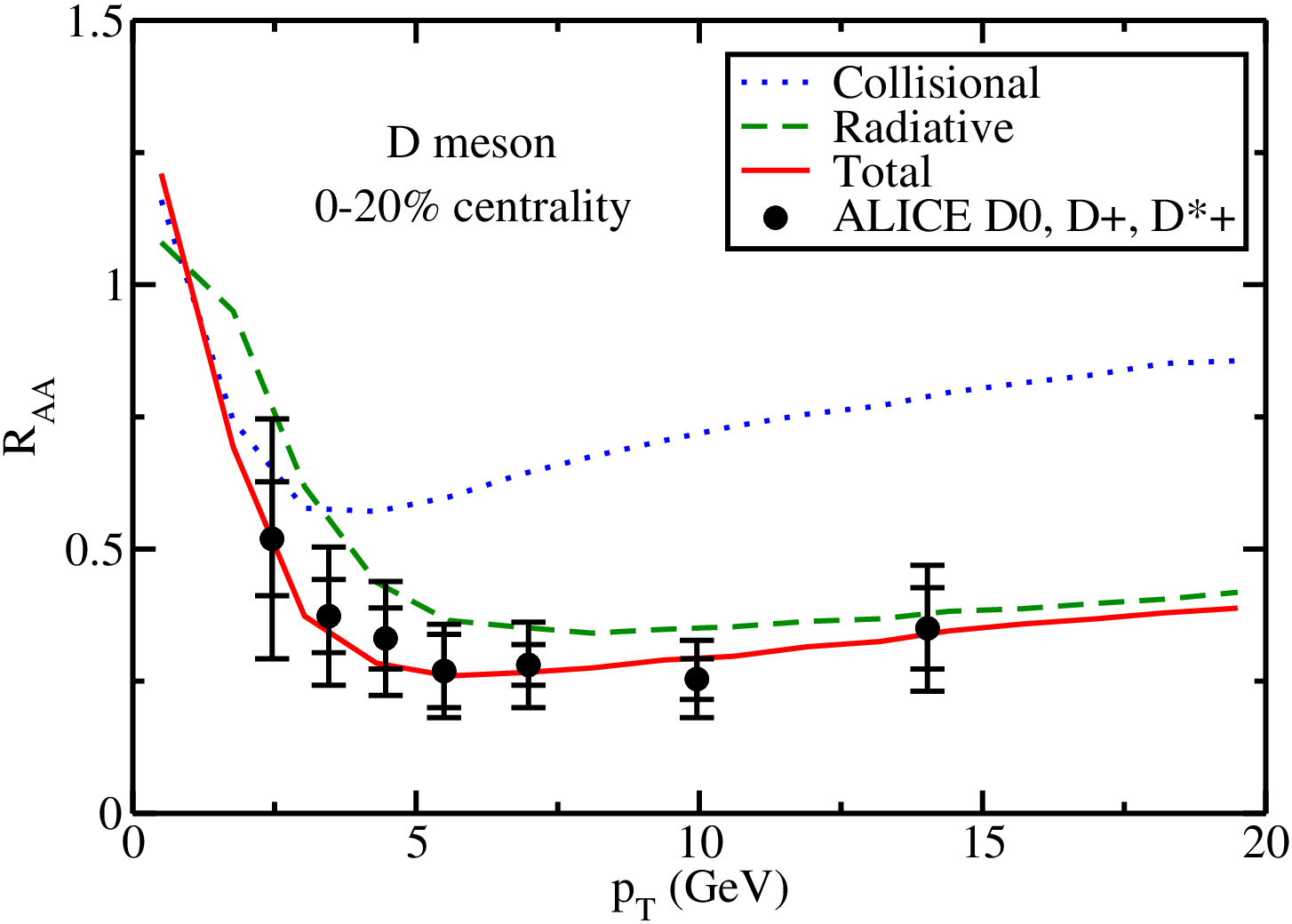}
\caption{\label{DRAA}(Color online) A comparison of D meson $R_{AA}$ between different mechanisms and the LHC data.}
\end{minipage}\hspace{0.06\textwidth}
\begin{minipage}{0.47\textwidth}
\includegraphics[width=1\textwidth,clip=]{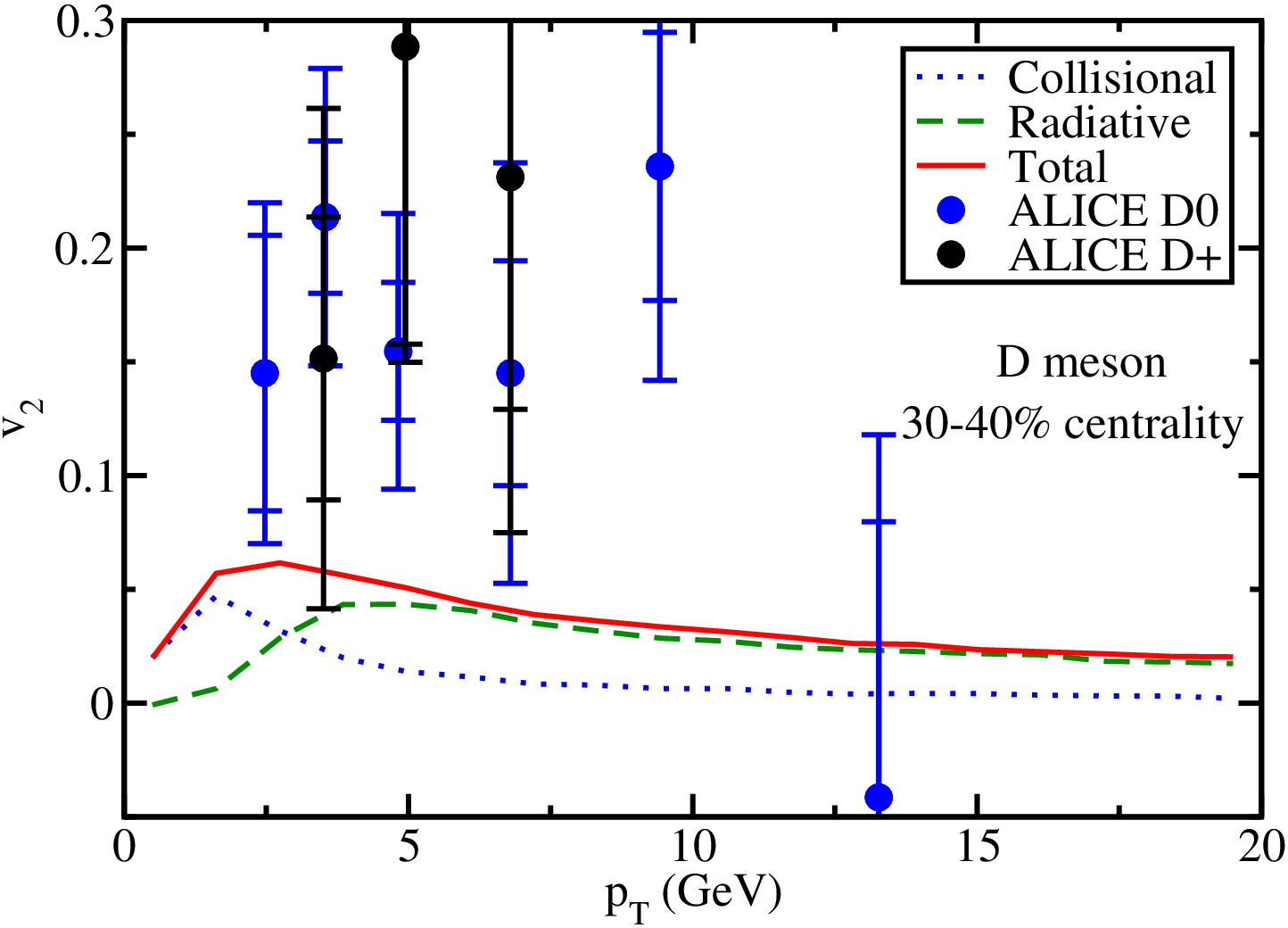}
\caption{\label{Dv2}(Color online) A comparison of D meson $v_2$ between between different mechanisms and the LHC data.}
\end{minipage}
\end{figure}

\begin{figure}[htb]
\begin{minipage}{0.47\textwidth}
\includegraphics[width=1\textwidth,clip=]{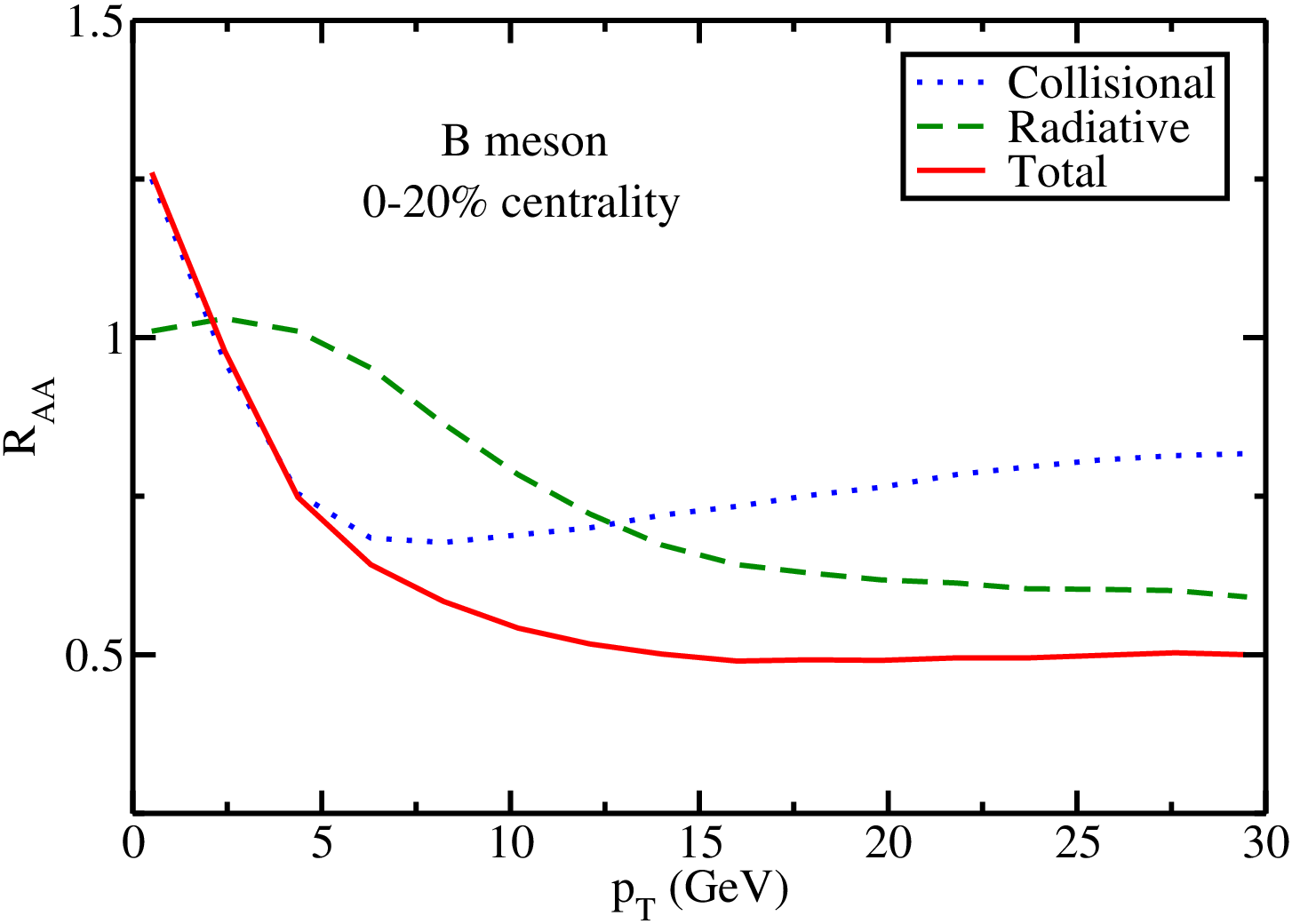}
\caption{\label{BRAA}(Color online) A prediction of B meson $R_{AA}$ with different energy loss mechanisms.}
\end{minipage}\hspace{0.06\textwidth}
\begin{minipage}{0.47\textwidth}
\includegraphics[width=1\textwidth,clip=]{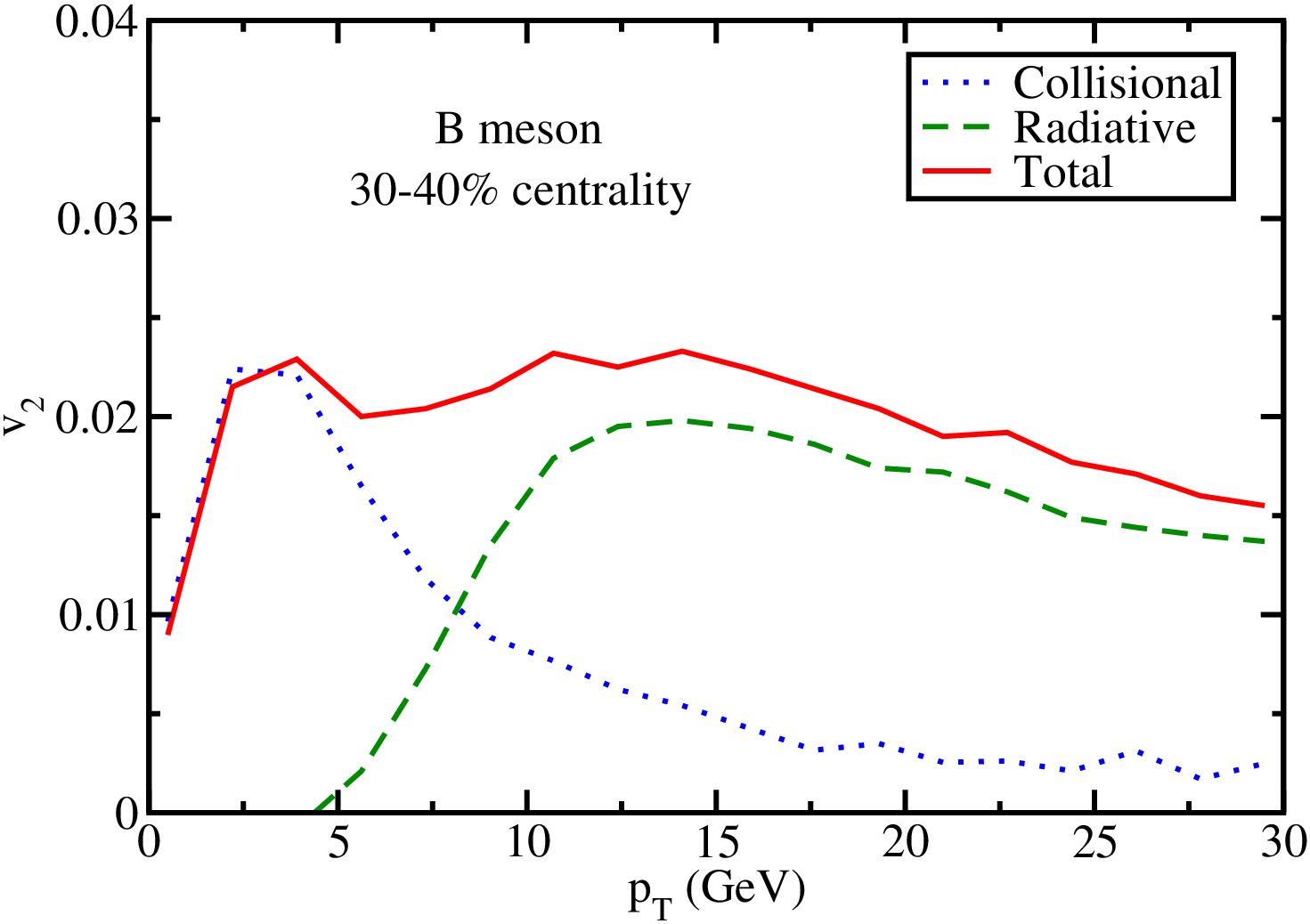}
\caption{\label{Bv2}(Color online) A prediction of B meson $v_2$ with different energy loss mechanisms.}
\end{minipage}
\end{figure}

Fig.\ref{DRAA} shows our calculation of D meson suppression compared with the LHC data. Neither collisional nor radiative energy loss alone are able to describe the data. However, our combination of the two mechanisms provide a good description of D meson $R_{AA}$ measured by ALICE. We also present our calculation of D meson $v_2$ in Fig.\ref{Dv2}, and our prediction for B meson suppression and flow in Fig.\ref{BRAA} and Fig.\ref{Bv2}. B mesons display similar behaviors as D mesons except that the crossing point between collisional and radiative dominating regions becomes much higher due to the larger mass of the bottom quark vs. the charm quark.

\section{Summary}

We have studied heavy quark evolution and energy loss inside a QGP medium in the framework of a Langevin equation. We have improved the current Langevin approach such that not only collisional, but also radiative energy loss is able to be incorporated by treating gluon radiation as an extra force term. Our calculation reveals a significant effect of the medium-induced gluon radiation on the heavy quark energy loss at LHC energies, which should no longer be neglected due to the ``dead-cone effect". And our combination of the two energy loss mechanisms provides a good description of the D meson suppression measured by the ALICE Collaboration.

\section*{Acknowledgments}

The (2+1)D viscous hydrodynamics model (VISH2+1) was developed by Huichao Song \cite{Song:2007fn,Song:2007ux} and recently modified by Zhi Qiu and Chun Shen for increased numerical stability and user friendliness. We here employed the code version and parameter tunings for Pb+Pb collisions at LHC energies that were previously used in Ref. \cite{Qiu:2011hf}. We thank the Ohio State University group for providing us with the corresponding initialization and hydrodynamic evolution codes.


\bibliographystyle{NPA}
\bibliography{SCrefs}

\end{document}